\documentclass[twocolumn,amsmath,amssymb,amssymb, aps, showpacs, pra]{revtex4}
\usepackage{fmtcount} 
\usepackage{graphicx}
\usepackage{dcolumn}
\usepackage{bm}
\usepackage{braket}
\begin{document}
\title{Coherent control of atomic excitation using off-resonant strong few-cycle pulses.}
\author{Pankaj K. Jha$^{1,*}$, Hichem Eleuch$^{1,2}$, Yuri V. Rostovtsev$^{1,3}$}
\affiliation{$^{1}$Institute for Quantum Science and Engineering and Department of Physics and Astronomy, Texas A\&M University, College Station, Texas 77843, USA\\
$^{2}$Institut National des Sciences Appliquees et de Technologie, 1080 Tunis, Tunisia\\
$^{3}$Department of Physics, University of North Texas, Denton, Texas 76203, USA}
\begin{abstract}
\noindent We study the dynamics of a two-level system driven by an off-resonant few-cycle pulse which has a phase jump $\phi$ at $t=t_{0}$, in contrast to many cycle pulses, under non rotating-wave approximation (NRWA). We give a closed form analytical solution for the evolution of the probability amplitude $|C_{a} (t)|$ for the upper level. Using the appropriate pulse parameters like phase-jump $\phi$,  jump time $t_{0}$, pulse width $\tau$,  frequency $\nu$ and Rabi frequency $\Omega_{0}$ the population transfer, after the pulse is gone, can be optimized and for the pulse considered here, enhancement of $10^{6}-10^{8}$ factor was obtained.
\noindent 
\end{abstract}
\maketitle
Modern pulsed lasers produce bursts of light that are both ultra-short and ultra-strong, exhibiting durations comparable to those of molecular vibrations, and electric fields rivaling those near an atomic nucleus \cite{H1}. Attosecond lasers, emitting pulses with only a few optical cycles per pulse \cite{H2}, hold the promise of controlling the phase difference between the carrier wave and its envelope\cite{H3}.

The interaction between strong, broadband electromagnetic fields and atoms, especially laser radiation that is tuned far from resonance, is of current interest. Short pulses can excite coherence on high-frequency transitions that may be used for efficient generation of XUV radiation \cite{H4,TLA2,TLA3}. Shaped pulses can enhance transient population of the excited state \cite{R1} or create optimal coherence in TLS \cite{R2}. Recently we have found a new analytical solution describing the dynamics of a two-level atom under the action of laser radiation with an arbitrary pulse shape and polarization \cite{H5}. Furthermore, we have studied two mechanisms of atomic excitation: multi-photon excitation, and breaking of adiabaticity \cite{H4}, and we have shown \cite{H6} that the latter can be the more efficient.

Interaction of such ultrashort pulses with a two-level atom under rotating-wave approximation does not give us the complete picture since the variation of the atomic polarization and population within the optical cycle is not slow. Thus we should not neglect the contribution of the counter-rotating terms in the Hamiltonian while studying few cycle pulses interaction with atomic systems \cite{FCP1,FCP2,FCP3,FCP4,FCP5,FCP6,hli10prl,jha10prl}. On the other hand if the fields are not too strong and the variation of the atomic polarization and population within the optical cycle is slow, RWA appears to be a good approximation.

In this Brief report we studied the interaction of few-cycle pulses, in contrast to many cycle pulses \cite{V1,V2,V3}, with two-level system. These pulse have a phase jump $\phi$ at $t=t_{0}$. Thus they can be characterized by the parameters peak Rabi frequency $\Omega_{0}$, pulse width $\tau$, carrier frequency $\nu$, phase jump $\phi$ and jump moment $t_{0}$ along with the pulse envelope ( which we have considered gaussian for the numerical simulation). We present an analytical solution for this problem. Using the appropriate characterizing parameters, the population transfer,  can be optimized and for the pulse considered here, enhancement of $10^{6}-10^{8}$ factor was obtained [see Fig. 5(b)].

The equation of motion for the probability amplitudes for the states $|a\rangle$ and $|b\rangle$ of a two-level atom (TLA) interacting with a classical field is given as~\cite{BK1}
\begin{subequations}\label{TLA1}
\begin{align}
\dot{C}_{a}&= i\frac{\wp{\cal E}(t)}{\hbar}\mbox{cos}(\nu t) e^{i\omega t}C_{b},\label{second}\\
\dot{C}_{b}&= i\frac{\wp^{*}{\cal E}(t)}{\hbar}\mbox{cos}(\nu t)e^{-i\omega t}C_{a},
\end{align}
\end{subequations}
\noindent where $\hbar\omega$ is the energy difference between two levels, $\wp$ is the atomic dipole moment; $E(t)={\cal E}(t)\mbox{cos}\nu t$. In the rotating-wave approximation (RWA) we replace $\mbox{cos}(\nu t)e^{\pm i\omega t}\rightarrow e^{\pm i\Delta t}/2$ where $\Delta=\omega-\nu$ \cite{FN1}, is detuning from resonance. Introducing $\Omega(t)=\wp{\cal E}(t)/\hbar$, Eq.(\ref{TLA1}) reduces to
\begin{subequations}\label{TLA2}
\begin{align}
\dot{C}_{a}&= i\frac{\Omega(t)}{2} e^{i\Delta t}C_{b},\label{second}\\
\dot{C}_{b}&= i\frac{\Omega^{*}(t)}{2}e^{-i\Delta t}C_{a},
\end{align}
\end{subequations}which have an integral of motion $|C_{a}|^{2} + |C_{b}|^{2} = 1$. If we define a function $f(t)=C_{a}(t)/C_{b}(t)$, Eq.(\ref{TLA2}) yields the following Riccati Equation
\begin{equation}\label{L3}
\dot{f}+i\frac{\Omega^{*}(t)}{2}e^{-i\Delta t}f^{2}-i\frac{\Omega(t)}{2}e^{i\Delta t}=0.
\end{equation}
Then $|C_{a}(t|=|f(t)| / \sqrt{1+|f(t)|^{2}}$. Alternatively, we can get a second order linear differential equation for $C_{a}(t)$, from Eq.(\ref{TLA2})
\begin{equation}\label{L4}
\ddot{C}_{a}(t)-\left[i\Delta + \frac{\dot{\Omega}(t)}{\Omega(t)}\right]\dot{C}_{a}(t)+\frac{|\Omega(t)|^{2}}{4}C_{a}(t)=0.
\end{equation}
In this paper we will work without RWA, hence the Riccati Eq.(\ref{L3}) takes the new form as
\begin{equation}\label{L5}
\dot{f}+i\Omega^{*}(t)\text{cos}(\nu t)e^{-i\omega t}f^{2}-i\Omega(t)\text{cos}(\nu t)e^{i\omega t}=0.
\end{equation}
\begin{figure}[t]
  \includegraphics[width=0.42\textwidth,height=3.4cm]{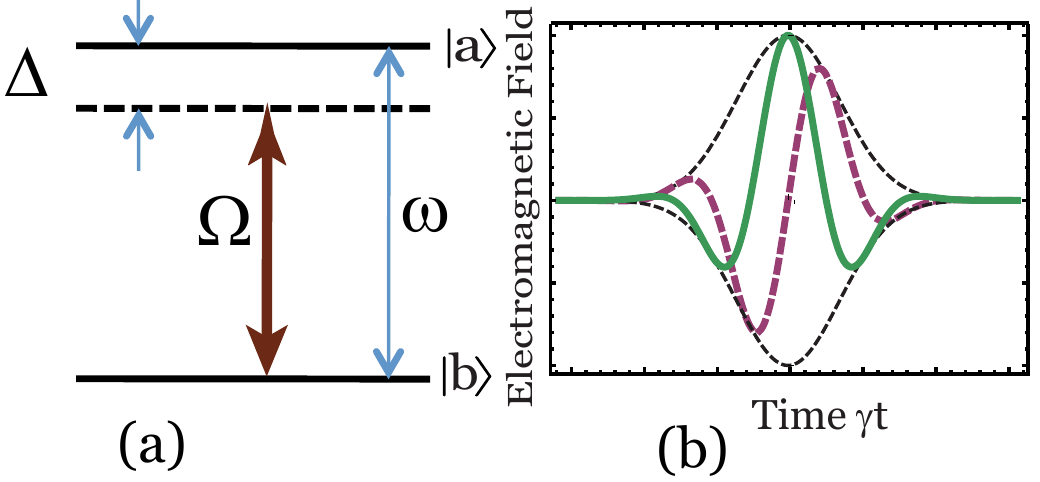}
  \caption{(Color Online) (a) Two-level atomic system, atomic transition frequency $\omega=\omega_{a}-\omega_{b}$, detuning $ \Delta = \omega -\nu$ and Rabi frequency $\Omega(t)=\wp{\cal E}(t)/\hbar$. (b) Few cycle sine (dashed line) and cosine (solid line) pulse with gaussian envelope.}
\end{figure}
The approximate solution for Eq.(\ref{L5}), in terms of the tip angle $\theta$ is given as \cite{H5}
\begin{equation}\label{L6}
\begin{split}
f(t)&=i\int_{-\infty}^{t}dt'\left\{\left[\frac{d\theta(t')}{dt'}-\theta^{2}(t')\frac{d\theta^{*}(t')}{dt'}\right]\right.\\
&\left. \times\text{exp}\left[2\int_{t'}^{t}\theta(t'')\dot{\theta}^{*}(t'')dt''\right] \right\},
\end{split}
\end{equation}
where the tip angle $\theta(t)$ has been defined as 
\begin{equation}\label{L7}
\theta(t)=\int_{-\infty}^{t}\Omega(t')\text{cos}(\nu t')e^{i\omega t'}dt'.
\end{equation}
From Eq.(\ref{L7}) we can obtain $|C_{a}(t)|=|f(t)|/\sqrt{1+|f(t)|^{2}}$. To see how well the approximate solution works, we have plotted the probability amplitude $|C_{a}(\infty)|$ for a complex pulse shape given by $\Omega(t)=\Omega_{0}[\text{sech}(\alpha t)+\text{sech}(\alpha t-3)]$ [see Fig 2]. Numerical simulation (dashed) and analytical solution (solid) shown in Figs. 2(a,b) are nearly identical.
\subsection{Pulse with phase jump}
In this section we will investigate the dynamics of a two-level atom subjected to few-cycle pulse with a phase jump at an arbitrary time $t=t_{0}$. Let us define the Rabi frequency $\Omega(t)$ for our model as 
\begin{equation}\label{L8}
\Omega(t)= 
\begin{cases} \Omega_{-}(t) & \text{if $t<t_{0}$,}
\\
\Omega_{+}(t) &\text{if $t\ge t_{0}$,}
\end{cases}
\end{equation}
where $\Omega_{+}(t)=e^{i\phi}\Omega_{-}(t)$ and $\phi$ is the phase jump introduced to the electromagnetic field at $t=t_{0}$. Equivalently the tip angle define by Eq.(\ref{L7}) takes the form
\begin{equation}\label{L9}
\theta(t)= 
\begin{cases} \theta_{-}(t) & \text{if $t<t_{0}$,}
\\
\theta_{+}(t) &\text{if $t\ge t_{0}$.}
\end{cases}
\end{equation}
\begin{figure}[t]
  \includegraphics[width=0.46\textwidth,height=3.0cm]{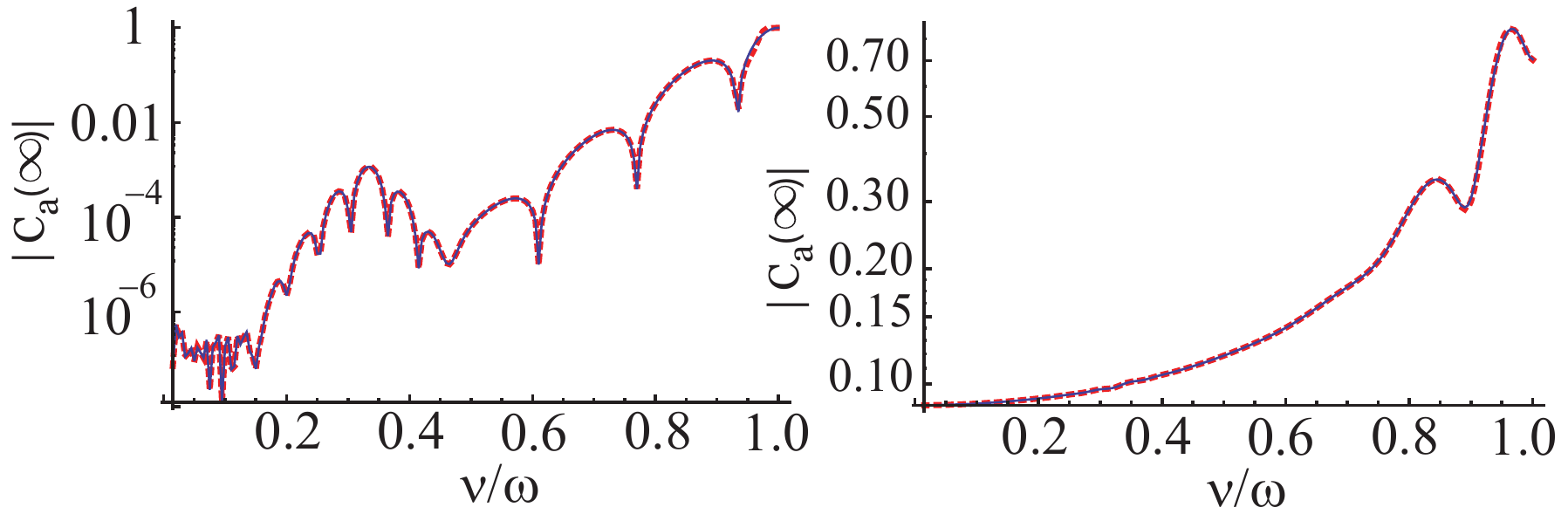}
  \caption{(Color Online) Population left in the upper level $|a\rangle$ after applying $\Omega(t)=\Omega_{0}[\text{sech}(\alpha t)+\text{sech}(\alpha t-3)]$ pulse as a function of the frequency $\nu/\omega$ obtained by numerical solution of Eq.(\ref{TLA1}) (dots) and using our approximate analytical result Eq.(\ref{L6}) (solid line). In calculations we take $\Omega_{0}=0.04 \omega$ and $\alpha=0.075 \omega$. In (a) $\phi=0$ while in (b) $\phi=\pi, t_{0}=0$}
\end{figure}
From the definition of the Rabi frequency Eq.(\ref{L8}), we can easily see that $\theta_{+}=e^{i\phi}\theta_{-}$.The time evolution of our system is divided into two regimes $(-\infty,t_{0})$ and $(t_{0},\infty)$. In both these regimes, the functional form of the solutions remains the same. We can write
\begin{equation}\label{L10}
f_{\phi}(t)= 
\begin{cases} f_{-}(t) & \text{if $t<t_{0}$,}
\\
f_{+}(t) &\text{if $t\ge t_{0}$.}
\end{cases}
\end{equation}
Eq.(\ref{L6}) is the solution for $\phi=0$ for the initial condition $f(-\infty)=0$. Using the same initial condition we can safely write 
\begin{equation}\label{L11}
f_{-}(t) =i\int_{-\infty }^{t}dt^{\prime }\Phi_{-}(t') \text{exp}\left[ 2\int_{t^{\prime }}^{t}\zeta_{-}(t'')dt^{\prime\prime }\right],
\end{equation}
where
\begin{subequations}\label{L12}
\begin{align}
\Phi_{-}(t')&=\left[ \frac{d\theta_{-}(t^{\prime })}{dt^{\prime }}-\theta^{2}_{-}(t^{\prime})\frac{d\theta_{-}^{\ast }(t^{\prime})}{dt^{\prime }}\right] \label{second},\\
\zeta_{-}(t'')&=\theta_{-}(t^{\prime \prime })\dot{\theta}_{-}^{\ast }(t^{\prime \prime }).
\end{align}
\end{subequations}
As the functional form of $f_{+}(t)$ and $f_{-}(t)$ are the same, we can write 
\begin{equation}\label{L13}
f_{+}(t) =i\int_{t_{0} }^{t}dt^{\prime }\Phi_{+}(t') \text{exp}\left[ 2\int_{t^{\prime }}^{t}\zeta_{+}(t'')dt^{\prime\prime }\right] +c,
\end{equation}
where $\Phi_{+}(t')=e^{i\phi}\Phi_{-}(t')$ and $\zeta_{+}(t')=\zeta_{-}(t')$
The constant $c$ can be obtained by demanding the continuity of $f_{\phi}(t)$ at $t=t_{0}$ which gives 
\begin{equation}\label{L14}
c=i\int_{-\infty }^{t_{0}}dt^{\prime }\Phi_{-}(t') \text{exp}\left[ 2\int_{t^{\prime }}^{t_{0}}\zeta_{-}(t'')dt^{\prime\prime }\right].
\end{equation}
\begin{figure}[t]
  \includegraphics[width=0.46\textwidth,height=6.2cm]{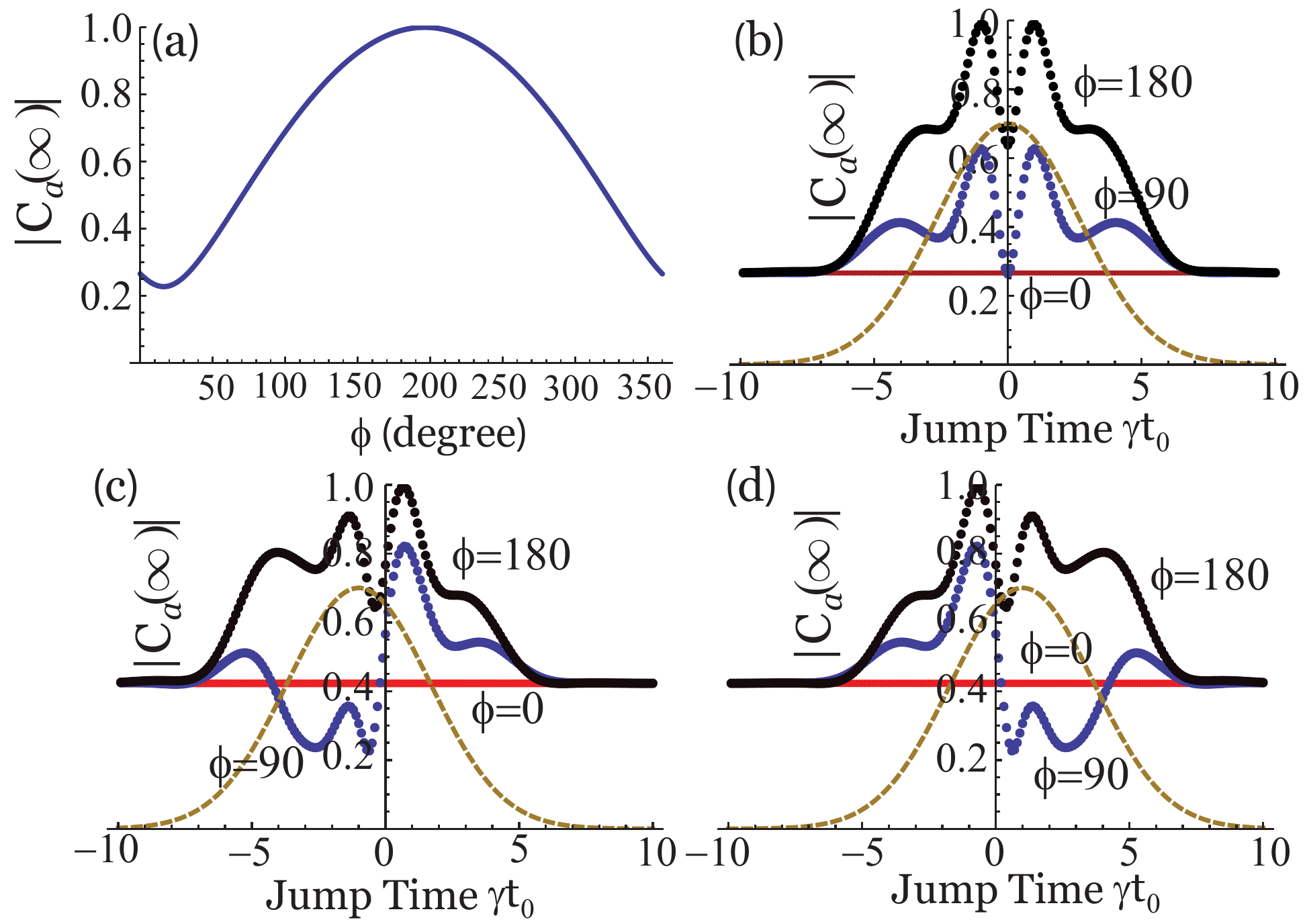}
    \caption{(Color Online) Effect of jump time $t_{0}$. (a) Here we have plotted the probability amplitude $|C_{a}(\infty)|$ against the phase jump $\phi$. Phase jump is introduced at the peak of the gaussian envelope. (b) The symmetric influence on the degree of excitation with respect to the position of $t_{0}$. The symmetric response is lost for shifted gaussian input pulse (c) and (d). For numerical calculations we chose $\Omega_{0}=0.875\omega$, $\nu=0.75\omega$, $\alpha=0.331 \omega$ and $\gamma=1.25 \omega$.  }
\end{figure}
\begin{figure}[b]
  \includegraphics[width=0.46\textwidth,height=6.5cm]{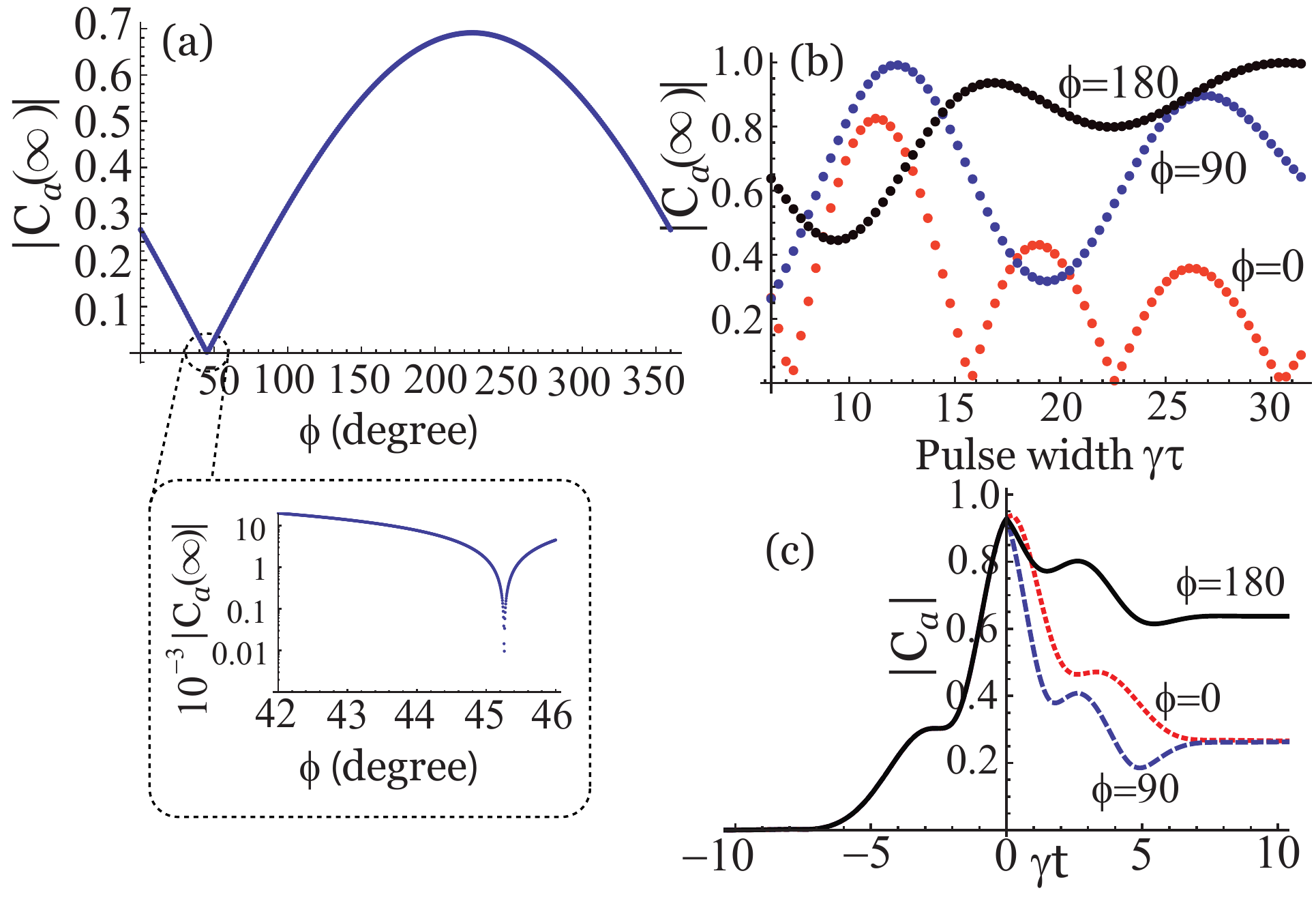}
  \caption{(Color Online) Effect of $\alpha$. (a) Probability amplitude $|C_{a}(\infty)|$ varies in the range from $10^{-5} \backsim  0.7$. (b) We have plotted $|C_{a}(\infty)|$ against normalized pulse width $\gamma \tau $ for fixed $\omega, \nu, \Omega_{0}$ and three combinations of the phase jump $\phi = 0,\pi/2, \pi$. (c) Shows the temporal evolution for the three combinations used in (b). For numerical simulation we chose $\Omega_{0}=0.875\omega$, $\nu=0.75\omega$, $\gamma=1.25 \omega$ and $\alpha=0.331 \omega$.  }
\end{figure}
\noindent Population transferred to the level $|a\rangle$ during the interaction is given as $|C_{a}(\infty)|^{2}=|f_{\phi}(\infty)|^{2}\big/(1+|f_{\phi}(\infty)|^{2})$. In order to study the effect of the phase jump $\phi $ let us define a relative change in the amplitude
\begin{equation}\label{L15}
r_{\phi }(t)=\left\vert \frac{f_{\phi}(t)-f(t)}{f(t)}\right\vert .
\end{equation}
Using Eq.(\ref{L11}), Eq.(\ref{L13}) and Eq.(\ref{L6}) we get,
\begin{equation}\label{L16}
r_{\phi }(t)= \left\vert\frac{(e^{i\phi }-1)\int_{t_{0}}^{t}dt^{\prime }\Phi_{-}(t') \text{exp}\left[ 2\int_{t^{\prime }}^{t}\zeta_{-}(t'')dt^{\prime\prime }\right]}{\int_{-\infty}^{t}dt^{\prime }\Phi_{-}(t') \text{exp}\left[ 2\int_{t^{\prime }}^{t}\zeta_{-}(t'')dt^{\prime\prime }\right]}\right\vert .
\end{equation}
The asymptotic value $r_{\phi}(\infty)$ can be obtained by $t \rightarrow \infty$ in Eq.(\ref{L16}). We can easily see from the Eq.(\ref{L16}), that $r_{\phi}(\infty)$ attains its maximum value for $\phi=\pi$.
\subsection{Effect of Pulse parameters: Numerical Analysis}
In this section we will discuss the effect of the pulse parameters like phase jump time $t_{0}$, pulse witdth $\tau$, detuning $\Delta$ and peak Rabi frequency $\Omega_{0}$ on the degree of excitation of the upper level $|a\rangle$. For the computational purpose we have considered a Gaussian pulse of the form $\Omega(t)=\Omega_{0}e^{-\alpha^{2}t^{2}}$ where $\alpha=2\sqrt{\text{ln}2}/\tau$ ($\tau$ is the FWHM of the pulse).
\begin{figure}[t]
  \includegraphics[width=0.46\textwidth,height=3.0cm]{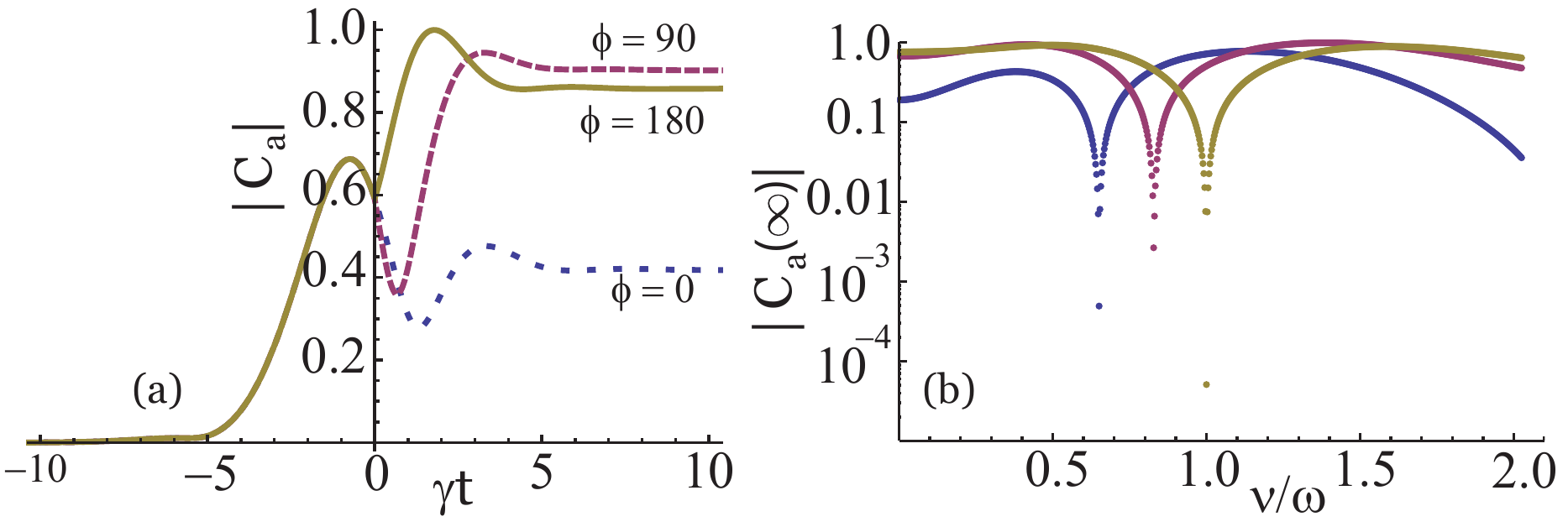}
  \caption{(Color Online) (a) Temporal behavior of $|C_{a}|$ for difference combination of $\phi$. (b) Plot of $|C_{a}(\infty)|$ against $\nu/\omega$. For numerical simulation we chose $\Omega_{0}=0.875\omega$, $t_{0}=0,  \gamma=1.25 \omega$ and $\alpha=0.331 \omega$.
  }
\end{figure}
\begin{figure}[b]
  \includegraphics[width=0.46\textwidth,height=3.2cm]{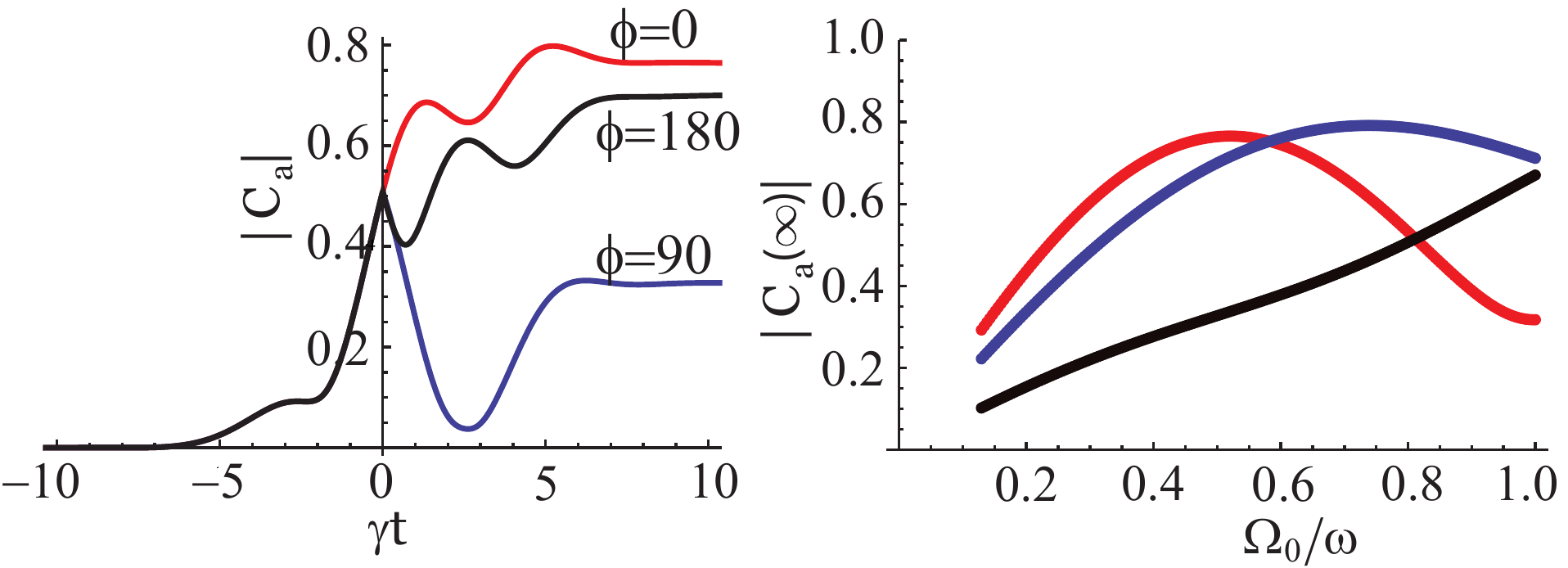}
  \caption{(Color Online) (a) Temporal behavior of $|C_{a}|$ for difference combination of $\phi$. (b) Plot of $|C_{a}(\infty)|$ against $\Omega_{0}$. For numerical simulation in (b), we chose a shifted gaussian pulse with $t_{s}=1$,  $\nu=0.75\omega$, $t_{0}=0,  \gamma=1.25 \omega$ and $\alpha=0.331 \omega$.  $\Omega_{0}=0.875\omega$ for Fig. (a).  }
\end{figure}

The main result showing the effect of relative position of $t_{0}$, with respect to the peak of the pulse, on the atomic excitation is shown in Fig. 3 and Fig. 7(a) where we have shown the dynamics of the two-level atom interacting with few-cycle pulse with a phase jump. In Fig. 3(a) we have one such scenario of $\phi=\pi/2$.  Here the phase jump is introduced in the field at the peak of the gaussian envelope i.e $t_{0}=0$ and plotted the probability amplitude $|C_{a}(\infty)|$ against the phase jump $\phi$. Interestingly the difference in the maximum and the minimum value corresponds to $\Delta \phi =\pi$. The symmetric nature of the atomic excitation is observed in Fig. 3(b) and the contour plot Fig. 7(a). With the shifted Gaussian pulse $\Omega=\Omega_{0}e^{-\alpha^{2}(t\pm t_{s})^{2}}$ [see Figs. 3(c) and 3(d)] the symmetry is lost. Also the effect of the phase jump becomes significant for $t_{0}$ within the FWHM of the pulse and gradually decreases when $t_{0}$ is close to the tail of the pulse. Identical response of the system, for $\gamma t_{0} \approx 10$, is observed for three combinations of the phase jump $\phi = 0,\pi/2, \pi$.
\begin{figure}[t]
  \includegraphics[width=0.46\textwidth,height=9.4cm]{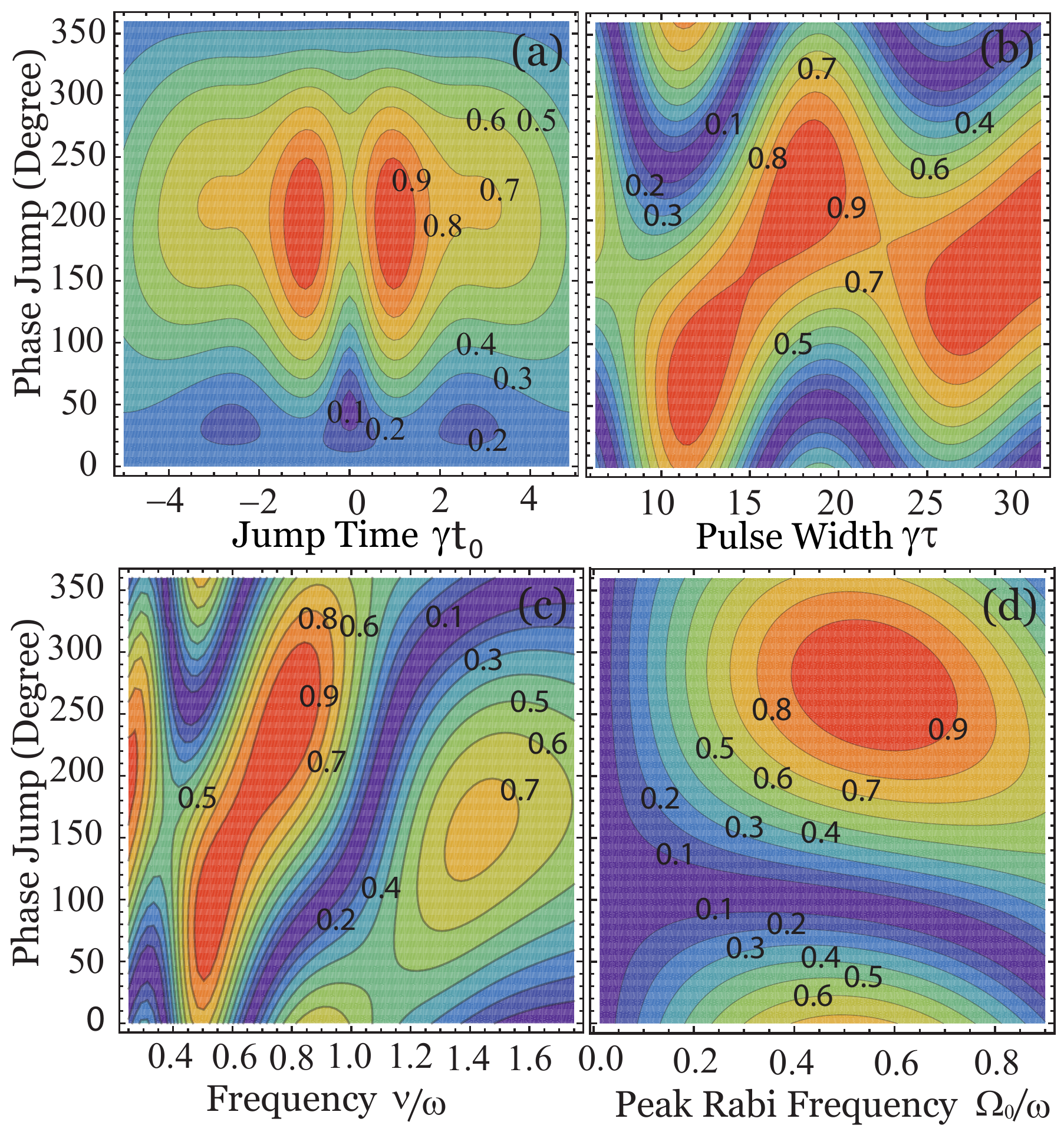}
  \caption{(Color online) Contour plot showing the effect of pulse parameters like $t_{0},\tau,\nu$ and $\Omega_{0}$ on the population left in the excited states $|a\rangle$ in (a), (b), (c) and (d) respectively. The influence of the phase jump time $t_{0}$ is symmetric with respect to $t_{0}$ is shown in (a). Parameters used are $\Omega_{0}=0.875\omega$, $\nu=0.75\omega$, $\gamma=1.25 \omega, , t_{0}=0$ and $\alpha=0.331 \omega$ as required appropriately. For (c) we used $\alpha=0.110 \omega$ }
\end{figure}

While investigating the effect of few-cycle pulses on atomic systems, the parameter $\alpha$ plays an important role for a given value of the carrier frequency $\nu$. It determines the number of cycles of the field in the pulse. The main results showing the effect of $\alpha$ or the pulse width $\tau$ is given in Fig. 4 and contour plot Fig. 7(b). If we look at the inset of Fig. 4(a) we see that the probability amplitude $|C_{a}(\infty)|$ varies in the range from $10^{-5} \backsim  0.7$.  In Fig 4(b) we have used three combination of phase jump $\phi$ ($\phi = 0,\pi/2, \pi$) to study the effect of $\alpha$ on the degree of excitation. For lower pulse width ($ 2 \le \gamma \tau \le 15)  \phi =\pi/2$ creates more excitation than $\phi=0$ or $\pi$.

In order to study the effect of detuning $\Delta$ we have plotted the response of the system in terms of $|C_{a}(\infty)|$ for the three combination of $\phi$. Fig 5(a) shows the temporal behavior while Fig. 5(b) gives the information about steady-state population. The probability amplitude $|C_{a}(\infty)|$ varies in the range from $ 4.4 \times 10^{-4} \backsim  0.4$ for $\phi=0$ and $ 5 \times 10^{-5} \backsim  0.9$ for $\phi=\pi$. When $|C_{a}(\infty)|$ is $\backsim 4.4 \times 10^{-4}$ for $\phi=0$ we have $|C_{a}(\infty)| \backsim 1$ for $\phi=\pi$, thus we have an enhancement of $10^{6} - 10^{8}$ factor in the population transfer by introducing a phase jump of $\pi$ at the peak of the envelope function.

The effect of the peak Rabi frequency $\Omega_{0}$ on the degree of excitation of the upper level in shown in Fig 6 and the contour plot Fig 7(d). While Fig 6(a) shows the temporal behavior of $|C_{a}|$ on the other hand Fig 6(b) gives the information about the population left in the upper level after the pulse is gone. We see that for some choice of $\Omega_{0}$ $\phi=0$ has the maximum effect while for some $\phi=\pi/2$ is dominant.

In conclusion, we have studied few-cycle pulses, with a phase jump $\phi$ at $t=t_{0}$, interacting with a two-level atom. This interaction is investigated without the rotating-wave approximation and we present an approximate solution for the probability amplitude $C_{a} (t)$ of the upper level.  The approximate solution not only works well with multi-cycle pulse~\cite{H5} but it is also in excellent agreement for few-cycles pulses [see Fig 2]. Using the appropriate pulse parameters $\phi$, $t_{0}$, $\alpha$ and $\Omega_{0}$ the population transfer, after the pulse is gone, can be optimized and for the pulse considered here, enhancement of $10^{6}-10^{8}$ factor was obtained [see Fig. 5(b)]

We thank M. O. Scully, L.V. Keldysh, and M. S. Zubairy for useful discussions and gratefully acknowledge the support from the NSF Grant No. EEC-0540832 (MIRTHE ERC), the Office of Naval Research (Grants No. N00014-09-1- 0888 and No. N0001408-1-0948), the Robert A. Welch Foundation (Award No. A-1261) and partial support from the CRDF. P.K.J would also like to acknowledge the Robert A. Welch Foundation and HEEP Foundation for financial support.


\begin{thebibliography}{99}
\bibitem{FN1}Here we use the convention that all frequencies are circular frequencies so that $\hbar\nu$ (not $h\nu$) is the photon energy. Some authors also use the definition of detuning $\Delta= \nu-\omega$.
\end{thebibliography}
\end{document}